\pdfoutput=1
\documentclass[aps,prl,reprint,showpacs,superscriptaddress]{revtex4-2}

\usepackage[pdftex]{graphicx}
\usepackage[pdftex]{epsfig}
\usepackage{epstopdf}
\usepackage{color}
\usepackage{amssymb,amsmath}
\usepackage{graphicx}
\usepackage{dcolumn}
\usepackage{multirow}
\usepackage[hypertexnames=false,colorlinks,urlcolor=blue,citecolor=blue]{hyperref}
\usepackage{siunitx}
\usepackage[dvipsnames]{xcolor}
\usepackage[capitalise]{cleveref}
\usepackage{float}

\usepackage{mathtools}

\newcommand{\papertitle}{Using a high-fidelity numerical model to infer the shape of a few-hole Ge quantum dot}

\begin{document}

\title{\papertitle}
\author{Mitchell Brickson}
\affiliation{Sandia National Laboratories, Albuquerque, NM 87185, USA}
\affiliation{Center for Quantum Information and Control (CQuIC), Department of Physics and Astronomy, University of New Mexico, Albuquerque, NM 87131, USA}
\author{N. Tobias Jacobson}
\affiliation{Sandia National Laboratories, Albuquerque, NM 87185, USA}
\author{Andrew J. Miller}
\affiliation{Sandia National Laboratories, Albuquerque, NM 87185, USA}
\author{Leon N. Maurer}
\affiliation{Sandia National Laboratories, Albuquerque, NM 87185, USA}
\author{Tzu-Ming Lu}
\affiliation{Sandia National Laboratories, Albuquerque, NM 87185, USA}
\author{Dwight R. Luhman}
\affiliation{Sandia National Laboratories, Albuquerque, NM 87185, USA}
\author{Andrew D. Baczewski}
\affiliation{Sandia National Laboratories, Albuquerque, NM 87185, USA}
\affiliation{Center for Quantum Information and Control (CQuIC), Department of Physics and Astronomy, University of New Mexico, Albuquerque, NM 87131, USA}

\begin{abstract}
The magnetic properties of hole quantum dots in Ge are sensitive to their shape due to the interplay between strong spin-orbit coupling and confinement.
We show that the split-off band, surrounding SiGe layers, and hole-hole interactions have a strong influence on calculations of the effective $g$ factor of a lithographic quantum dot in a Ge/SiGe heterostructure.
Comparing predictions from a model including these effects to raw magnetospectroscopy data, we apply maximum-likelihood estimation to infer the shape of a quantum dot with up to four holes.
We expect that methods like this will be useful in assessing qubit-to-qubit variability critical to further scaling quantum computing technologies based on spins in semiconductors.
\end{abstract}

\maketitle

\textit{Introduction.---} Spin qubits in lithographic quantum dots (QDs)~\cite{burkard2023semiconductor} are strong candidates for realizing fault-tolerant quantum computation, with recent demonstrations of high-fidelity operations~\cite{noiri2022fast,xue2022quantum,weinstein2023universal} and the promise of leveraging the same fabrication technologies as classical devices with billions of transistors to scale up~\cite{zwerver2022qubits,neyens2024probing}.
Among QD technologies, hole spins in Ge have progressed rapidly~\cite{watzinger2018germanium,hendrickx2020fast,hendrickx2020single,hendrickx2021four,jirovec2021singlet,scappucci2021germanium,borsoi2023shared,zhang2023universal}. 
They experience strong intrinsic spin-orbit coupling (SOC) that enables all-electrical control~\cite{moriya2014cubic}, benefit from small effective masses that facilitate tunable exchange coupling~\cite{laroche2016magneto}, can be integrated with superconductors~\cite{hendrickx2018gate}, and have decreased sensitivity to hyperfine noise-- with or without isotopic purification~\cite{bosco2021fully}.
Access to high-quality Ge/SiGe quantum wells~\cite{sammak2019shallow,stehouwer2023germanium} has led to the successful demonstration of multiple 4-qubit devices~\cite{hendrickx2021four,zhang2023universal}, as well as a 16-dot array~\cite{borsoi2023shared}.
However, the complexity of SOC and disorder in the host material can lead to significant variability between QDs, even within the same device~\cite{hendrickx2021four,jirovec2021singlet,hendrickx2023sweet}.
Understanding the relationships between microscopic device physics, QD properties, and qubit performance will be essential to realizing large arrays of QDs with little variation from qubit to qubit~\cite{pena2024modeling}.

\begin{figure*}[ht]
\includegraphics[width=0.85\textwidth]{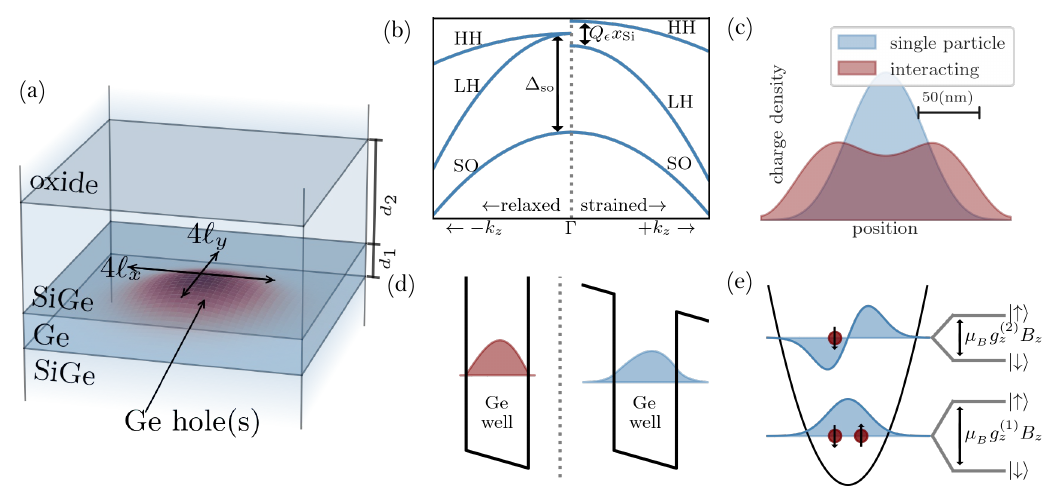}
\caption{
Details of our model of the Ge-hole QD first reported in Ref.~\cite{miller2022effective}.
(a) The QD is inside of a $d_1=\SI{20}{\nano\meter}$ Ge well. 
The well is on top of a Si$_{0.25}$Ge$_{0.75}$ half space and below a $d_2=\SI{62}{\nano\meter}$ capping layer of Si$_{0.25}$Ge$_{0.75}$ under an SiO$_2$ half space.
The confining potential (Eq.~\ref{eq:confining_potential}) traps holes with a ground-state single-particle width of $4\ell_\mu$ along the $\mu \in \lbrace x,y \rbrace$ axes~\cite{SMref}.
(b) Our LK Hamiltonian accounts for the valence band structure of the Ge well, in which strain splits the HH and LH bands by $Q_{\epsilon}x_\mathrm{Si}$=\SI{75}{\milli\electronvolt} (for $x_\mathrm{Si}=0.25$) and includes the SO band which is separated by $\Delta_{SO}$=\SI{296}{\milli\electronvolt}~\cite{SMref}.
(c) The Coulomb interaction between the holes is included using configuration interaction. 
Exemplary 1D slices of hole charge densities for single-particle and interacting models demonstrate that this leads to a significant change in the size and shape of the orbitals for a fixed confining potential.
(d) Accounting for the finite band offset between the Ge well and Si$_{0.25}$Ge$_{0.75}$ layers (right) leads to more delocalization of the hole orbitals relative to a model in which that band offset is infinite (left).
(e) Due to the differential effect of SOC on orbitals with different symmetry, the effective $g$ factor depends on whether the holes reside in the first (1) or second (2) shells.
\label{fig:overview}
}
\end{figure*}

These relationships are particularly nontrivial for holes in Ge due to strong SOC.
The light-hole/heavy-hole (LH/HH) degeneracy of the valence band maximum of bulk Ge is at the root of this complexity.
In SiGe/Ge heterostructures, significant strain in the Ge quantum well lifts this degeneracy (see Fig.~\ref{fig:overview}(b)) and isolates an effective spin-1/2 qubit subsystem in the HH subspace of a single spin-3/2 hole~\cite{venitucci2018electrical,hetenyi2020exchange,terrazos2021theory}.
Nevertheless, the HH and LH states remain coupled and qubit models necessarily account for at least these four bands-- either directly or perturbatively.
In fact, this coupling facilitates single-hole qubit control through electric-dipole spin resonance~\cite{crippa2018electrical,watzinger2018germanium}.
Even short of integrated qubit properties like Rabi frequencies or relaxation/dephasing times, seemingly simple magnetic properties like the effective $g$ factor are highly sensitive to the QD shape.

Other authors have measured the effective $g$ factor for magnetic fields applied in specific directions~\cite{watzinger2018germanium,hendrickx2020single,jirovec2021singlet}, and some have even measured the full $g$ tensors of single-hole QDs in Ge~\cite{hendrickx2023sweet}.
The sensitivity of the $g$ tensor to externally applied fields is directly tied to qubit operation.
For single-spin encodings the derivative of the $g$ tensor with respect to a controlling gate voltage matches measurements of the Rabi frequency~\cite{hendrickx2023sweet}, and the $g$ tensor differences between QDs provide the effective magnetic field gradient essential to implementing singlet-triplet encodings~\cite{jirovec2021singlet,mutter2021all,jirovec2022dynamics}.
The collective data from all of these experiments indicate that the $g$ factor can vary significantly from one QD to another, and can even change for an individual QD with changes in its confining potential~\cite{wang2022modelling,hendrickx2023sweet}.

Rationalizing this sensitivity with theory and modeling is challenged by the complexity of the physics in Ge hole QDs.
For example, data constraining the out-of-plane $g$ factor in a few-hole QD were reported in Ref.~\cite{miller2022effective}, with a 4-band non-interacting LK model~\cite{luttinger1955motion,luttinger1956quantum,winkler2003spin} achieving good agreement with experiment at 1- and 2-hole fillings, but poor agreement at 3-hole and higher fillings. 
This Letter first examines the impact of significantly more detailed numerical models of Ge-hole QDs, incorporating the split-off (SO) band, the effects of the surrounding SiGe layers, and hole-hole interactions dressed by the surrounding dielectric media.
We show that predictions of the $g$ factor can vary by a factor of nearly 3 among plausible but increasingly rich models.
Nevertheless, our results indicate that quantitative agreement between theory and experiment can be achieved at the level of raw magnetospectroscopy data, capturing both the QD charging energy and its magnetic properties.
In fact, thanks to the sensitivity of the properties of Ge hole QDs to their shape, we are able use maximum likelihood estimation (MLE) on these calculations to infer the shape of a QD in each of three fillings.
This is inspired in part by work on inferring the shapes of single-electron GaAs QDs from orbital energies~\cite{camenzind2019spectroscopy}.
Our approach is statistical in nature and captures the change in shape across multiple fillings.
We expect that such highly accurate models and techniques will be vital to assessing qubit-to-qubit variability as semiconductor qubit technologies continue to scale up in QD count.

\textit{Methods.---} We model the electrostatic potential defining the QD as
\begin{equation}
    V_{QD} = \frac{\left(\gamma_1+\gamma_2\right)\hbar^2}{2m_e}\left(\frac{x^2}{\ell_x^4}+\frac{y^2}{\ell_y^4}\right)+E_{z}z, \label{eq:confining_potential}
\end{equation}
where $\gamma_1$ and $\gamma_2$ are two of the three Luttinger parameters~\cite{luttinger1955motion,luttinger1956quantum,winkler2003spin}, $m_e$ is the rest mass of an electron, $E_z$ is the applied vertical electric field, and $\ell_x$ and $\ell_y$ are the lengths associated with harmonic confinement along the $x$ and $y$ axes, i.e., the shape of the QD.
$V_{QD}$ is used as the external potential in an LK Hamiltonian, the spectrum of which we compare to experimental magnetospectroscopy data to infer the values of $\ell_x$ and $\ell_y$. 
While $V_{QD}$ takes a simple functional form, other details of the LK Hamiltonian are relatively complicated and illustrated in Fig.~\ref{fig:overview}.
Simultaneously accounting for 4-6 bands, material discontinuities and attendant valence band offsets, and hole-hole interactions requires that we use a numerical method to analyze this model.
Specifically, we use an interior penalty discontinuous Galerkin (DG) discretization of the Hamiltonian in a basis of elemental-local higher order polynomials supported on a hexahedral mesh~\cite{hesthaven2007nodal,SMref}.

We compare a 4-band (HH/LH) LK Hamiltonian with a 6-band extension including the SO band.
Previous models of Ge hole QDs have often ignored this band because it is $\Delta_{SO}$=\SI{296}{\milli\electronvolt} away from the HH/LH subspace in bulk~\cite{winkler2003spin}, which is seemingly large relative to typical orbital splittings of $1-\SI{10}{\milli\electronvolt}$.
However, SOC differentially mixes \emph{both} the HH and LH bands with the SO bands, with a strength that is enhanced by the delocalization in momentum space created by strong confinement.
The HH and LH bands are also themselves coupled and this cumulative influence is further enhanced by strain, which augments the off-diagonal LH/SO coupling by as much as $\approx\!\SI{45}{\milli\electronvolt}$ for materials with 30\% Si in the alloy, i.e., $\approx $15$\%$ of $\Delta_{SO}$.
Because the effective $g$ factor is determined by the differential impact of the applied magnetic field on states that are mutually coupled and repelling one another, we expect that magnetospectroscopy may resolve the subtle differences in predictions between 4- and 6-band models.
We also expect this sensitivity to be greater for the higher orbital states occupied in few-hole dots, which are more anisotropic and have an extent in momentum space that increases with the orbital quantum number.

Another variable in our model is the level of detail used to describe the SiGe layers surrounding the Ge well in the LK Hamiltonian.
The valence band offset between Ge and SiGe can be as much as $\approx\!\SI{100}{\milli\electronvolt}$~\cite{schaffler1997high,terrazos2021theory}, and this is responsible for the vertical confinement of the QD.
Because this offset is also large relative to typical orbital splittings it is frequently approximated as infinite~\cite{secchi2021interacting,wang2021optimal}.
However, a recent analysis suggests that accounting for the penetration of the QD's orbitals into the surrounding alloy layers can significantly impact predictions of the effective $g$ factor~\cite{wang2022modelling}.
This is largely due to the influence of the LH bands, which are lighter along the $z$ axis and thus penetrate the SiGe layers more than the HH bands.
This is compounded by strain in the Ge, which pushes the LH band higher in energy and closer to the valence band edge in SiGe, and further compounded by the absence of strain in the SiGe layer, which enhances LH/HH mixing.
Thus, we consider both infinite and finite confining potentials, including the discontinuous Luttinger parameters and image charge effects due to the different dielectric constants in Ge and Si$_{0.25}$Ge$_{0.75}$~\cite{SMref}.


The Coulomb interaction between holes impacts not only the energy gaps between states of different occupancies, but the shapes of the orbitals at higher filling.
We include these effects using the full configuration interaction (CI) method~\cite{SMref}.
Using CI we calculate not only the spectrum of $N$-particle states for fixed $N$, but the energy differences between different occupancies that are probed in magnetospectroscopy.
Rather than comparing our model's predictions to effective $g$ factors, as in prior work~\cite{miller2022effective}, this allows us to compare directly with raw measurement data.
This can be particularly useful when it is difficult to unambiguously determine these slopes or even the associated spin-filling patterns.

Next, we apply MLE to predictions from our most detailed model, which includes 6 bands, the SiGe layers, and hole-hole interactions, to infer $\ell_x$ and $\ell_y$ in Eq.~\ref{eq:confining_potential}.
We note that parameters describing higher order anisotropy and anharmonicity could be inferred for a model of $V_{QD}$ with more parameters, though such an analysis would require rigorous model selection to avoid overfitting and we leave such an analysis to future work.
One could also account for measurements of other quantitites (e.g., Rabi frequencies) in the MLE to better constrain the inference, but this is also left to future work.

\begin{figure}[ht]
\includegraphics[width=0.9\columnwidth]{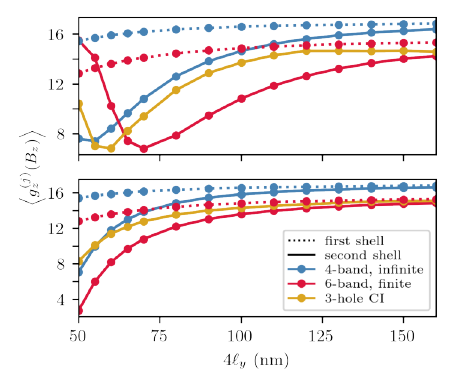}
\caption{
The effective out-of-plane $g$ factor for holes in the second shell as a function of the oscillator length along the $y$ axis, computed using three increasingly sophisticated LK Hamiltonians with confining potentials corresponding to different amounts of anisotropy.
The anisotropy strength is determined by the difference of the angular frequencies in $x$ and $y$, such that $\ell_x$ changes with $\ell_y$ and  $\omega_x-\omega_y=0$ is no anisotropy (top) while $\omega_x-\omega_y=\SI{0.5}{\milli\electronvolt}/\hbar$ is strong anisotropy (bottom).
The least sophisticated LK Hamiltonian (blue) only accounts for the LH and HH bands while approximating the confinement in $z$ as infinite and neglecting hole-hole interactions.
The next most sophisticated LK Hamiltonian (red) adds the SO band and accounts for finite confinement, while the most sophisticated LK Hamiltonian (yellow) includes these and hole-hole interactions.
\label{fig:g-factor_comparison}
}
\end{figure}

\textit{Results.---} While a simple 4-band infinite confinement model has been shown to agree reasonably well with measurements of the first-shell $g$ factor, this is not the case for the second-shell $g$ factor~\cite{miller2022effective}.
Thus we first consider the sensitivities of the latter to model complexity in Fig.~\ref{fig:g-factor_comparison}.
Our calculations involve averaging the slope of the energy difference between Zeeman-split 3-hole eigenstates as a function of the $z$-directed magnetic field strength, between 0 and 200 mT.
The filling of these states is such that the first two holes are paired in a singlet-like state and the spin of the remaining hole is unpaired.
We also consider a range of plausible QD shapes (values of $\ell_x$ and $\ell_y$) as another axis along which sensitivity can be assessed.
We find that the second-shell $g$-factor predictions vary by a factor of nearly 3 among the models considered.
Even for the first-shell, we find that a more detailed 6-band model reduces the predicted $g$-factor by $\approx$25\%.

One striking feature of our predictions is non-monotonicity in $g_z^{(2)}$ as a function of $\ell_x$ and $\ell_y$.
This is a consequence of SOC particular to non-$s$-like orbitals, with the relevant orbitals being $p$-like in our calculations.
SOC lowers the energy of states with opposing spin and orbital angular momentum, such that the second shell states have both opposite spin and opposite orbital angular momenta at small magnetic fields~\cite{SMref}.
This is overpowered at higher magnetic fields, where there is a crossover to a doublet in which these states have the same orbital angular momentum but different spins.
The field strength at which this crossover occurs is sensitive to model details, being impacted by both the zero-field splitting and the slopes of the energy levels at non-zero field.

\begin{figure}[ht]
\includegraphics[width=0.9\columnwidth]{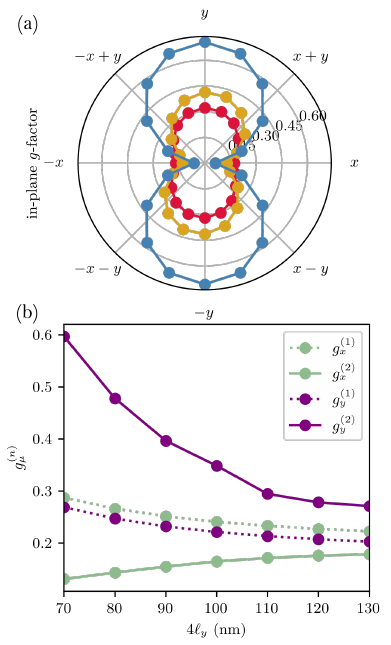}
\caption{
(a) In-plane second-shell $g$ factors as a function of the in-plane magnetic field angle.
This is a prediction for the potential selected from the MLE analysis (see Fig.~\ref{fig:mle_results}) with parameters $4\ell_y=\SI{110}{\nano\meter}$, $\hbar(\omega_x-\omega_y)=\SI{0.25}{\milli\electronvolt}$, and $E_z=\SI{1.5}{\volt/\micro\meter}$.
(b) In-plane $g$ factors for the first two shells as a function of $\ell_y$ for a fixed anisotropy in $V_{QD}$ of $\hbar(\omega_x-\omega_y)=\SI{0.25}{\milli\electronvolt}$.
Smaller dots tend to have stronger in-plane $g$-factor anisotropy, especially for the second shell.
\label{fig:in-plane_results}
}
\end{figure}

We also consider the sensitivities of predictions for the in-plane $g$ factors for first- and second-shell holes in Fig.~\ref{fig:in-plane_results}.
Across all models, the in-plane $g$ factors are smaller than the out-of-plane $g$ factors by about an order of magnitude, with the precise values decreasing with model detail.
The values predicted by our most detailed model (around $0.2$ to $0.3$) are consistent with experiment~\cite{hendrickx2020single,hendrickx2023sweet}.
Our calculations generally indicate significant anisotropy for second-shell holes due to both the anisotropy of $V_{QD}$ and the $p$-like character of the relevant orbitals.
This effect is stronger for smaller QDs than larger ones due to the increased extent in momentum space.
Finally, we note that the extrema of the in-plane $g$ factor are only nearly aligned with the $x$- and $y$-axes that define the harmonic oscillator axes.
This is due to the lack of cylindrical symmetry of the LK Hamiltonian about the $z$-axis and the fact that the $x$ and $y$ used here are not aligned with the crystal axes but instead with $x\rightarrow[110]$ and $y\rightarrow[\bar{1}10]$.

\begin{figure*}[!htb]
\includegraphics[width=0.9\textwidth]{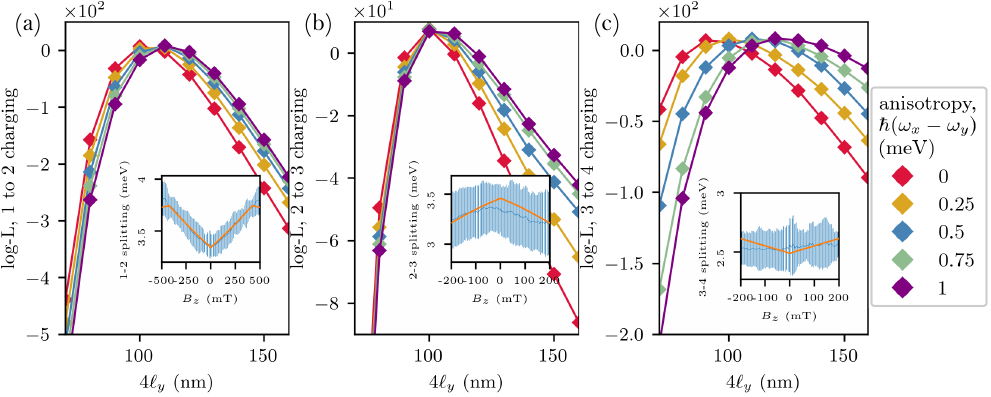}
\caption{
Log-likelihoods for the three different pairs of magnetospectroscopy lines as a function of the QD model's oscillator strengths, $\ell_x$ and $\ell_y$.
All predictions make use of the most sophisticated LK model under consideration, with 6 bands, finite confinement, and hole-hole interactions.
(Insets) the experimental magnetospectroscopy data (blue, with error bars) and the theoretical predictions from our model (orange) that are most consistent with the data.
\label{fig:mle_results}
}
\end{figure*}

This sensitivity suggests the possibility of inferring the potential from magnetospectroscopy data, so long as reasonable agreement can be achieved with predictions of both the effective $g$ factor and the charging energies.
Thus we apply MLE to raw data from Ref.~\cite{miller2022effective}, using a set of predictions from the 6-band model with finite confinement and a full configuration interaction treatment of hole-hole couplings.
The results of this analysis are shown in Fig.~\ref{fig:mle_results}, from which we infer the width and anisotropy of the harmonic potential most consistent with the data.
The predicted width from the 1- to 2-hole lines is $\SI{110}{nm}$ with a potential anisotropy of $\SI{0.75}{\milli\electronvolt}$, 2-3 is most consistent with a width of $\SI{100}{nm}$ and an anisotropy of $\SI{0.25}{\milli\electronvolt}$, and 3-4 is most consistent with a width of $\SI{120}{nm}$ and an anisotropy of $\SI{1}{\milli\electronvolt}$.
For the 1-2 and 2-3 charging energies, the anisotropy has very little effect on the optimal value of $\ell_y$ due to the $s$-like character of the relevant orbitals.
However, the 3-4 charging energies involve anisotropic $p$-like orbitals and thus show stronger variation of the optimal value of $\ell_y$ with changes in $\ell_x$.
The parameters producing the maximum likelihood indicate that the QD being is wider along $y$ than $x$, which is consistent with the device's gate geometry~\cite{miller2022effective}.

\textit{Conclusion.---} The magnetic properties of few-hole Ge-QDs are determined by the interplay between SOC, confinement, and hole-hole interactions.
We have explored how models of various levels of sophistication capture the impact of these physical mechanisms on effective hole g-factors.
Our richest model, which includes the split-off band, realistic confinement, discontinuities in the materials' parameters, and a configuration interaction treatment of hole-hole interactions, achieves good agreement with raw magnetospectroscopy data.
The level of agreement is sufficient to use MLE to infer the shape of a Ge-hole QD across three different fillings.
While the present analysis is restricted to a single QD in Ge, it could easily be extended to multiple QDs or different carriers and materials (e.g., electrons in Si).
Future work will consider cross-validation of inferred parameters among measurements of different properties, model selection to consider more sophisticated shapes, and an analysis of the impacts of disorder.

\begin{acknowledgments}
We gratefully acknowledge conversations with Susan Atlas, Quinn Campbell, Ivan Deutsch, David Dunlap, Malick Gaye, Shashank Misra, Alexandra Olmstead, and Vanita Srinivasa.
M.B., A.J.M., T.-M.L., D.R.L., and A.D.B. were partially supported by the Laboratory Directed Research and Development program at Sandia National Laboratories.
M.B. and A.D.B. were partially supported by the U.S. Department of Energy, Office of Science, National Quantum Information Science Research Centers, Quantum Systems Accelerator.
This work was performed, in part, at the Center for Integrated Nanotechnologies, an Office of Science User Facility operated for the U.S. Department of Energy (DOE) Office of Science.
This article has been authored by an employee of National Technology \& Engineering Solutions of Sandia, LLC under Contract No. DE-NA0003525 with the U.S. Department of Energy (DOE).
The employee owns all right, title and interest in and to the article and is solely responsible for its contents. 
The United States Government retains and the publisher, by accepting the article for publication, acknowledges that the United States Government retains a non-exclusive, paid-up, irrevocable, world-wide license to publish or reproduce the published form of this article or allow others to do so, for United States Government purposes. 
The DOE will provide public access to these results of federally sponsored research in accordance with the DOE Public Access Plan https://www.energy.gov/downloads/doe-public-access-plan.
\end{acknowledgments}

\bibliography{references}

\clearpage
\widetext
\begin{center}
\textbf{\large Supplemental Materials: \papertitle}
\end{center}

\setcounter{section}{0}
\setcounter{page}{1}
\setcounter{secnumdepth}{2}
\appendix
\makeatletter

The Supplemental Materials elaborate on certain details of of the central results in the main manuscript.
\begin{itemize}
    \item Appendix~\ref{app:extended_LK} provides details of the extended Luttinger-Kohn (LK) Hamiltonian used throughout.
    \item Appendix~\ref{app:numerical_methods} describes the numerical method used to discretize and solve our LK Hamiltonian.
    \item Appendix~\ref{app:mle_details} outlines the maximum-likelihood estimation (MLE) procedure used to infer quantum dot shapes.
    \item Appendix~\ref{app:zero-field} explains the origin of the non-monotonic dependence of $g^{(2)}_z(B_z)$ on $4\ell_y$, evident in Fig.~\ref{fig:g-factor_comparison}.
\end{itemize}

\section{Extended Luttinger-Kohn Hamiltonian}
\label{app:extended_LK}

\renewcommand{\theequation}{A\arabic{equation}}
\renewcommand{\thefigure}{A\arabic{figure}}
\setcounter{figure}{0}


\subsection{Definition}
\label{app:definition_LK}
One of our central results is the sensitivity of the magnetic properties of hole quantum dots to certain details of the LK Hamiltonian that have often gone ignored, with some exceptions.
While prior analyses have largely focused on 2- or 4-band models with complete confinement to the Ge quantum well, we report that a 6-band treatment that accounts for incomplete confinement and discontinuities in the material parameters is seemingly essential to achieve good agreement with the raw magnetospectroscopy data.
The ultimate form of this Hamiltonian~\cite{luttinger1955motion,luttinger1956quantum}, including Bir-Pikus strain effects~\cite{winkler2003spin}, is  
\begin{equation}
\label{eq:LK-Hamiltonian}
\hat{H}_\mathrm{LK}=\begin{bmatrix}
\hat{P}+\hat{Q}&-\hat{S}&R&0&\frac{1}{\sqrt{2}}\hat{S}&-\sqrt{2}\hat{R}\\
-\hat{S}^\dagger&\hat{P}-\hat{Q}&0&\hat{R}&\sqrt{2}\hat{Q}&-\sqrt{\frac{3}{2}}\hat{S}\\
\hat{R}^\dagger&0&\hat{P}-\hat{Q}&\hat{S}&-\sqrt{\frac{3}{2}}\hat{S}^\dagger&-\sqrt{2}\hat{Q}\\
0&\hat{R}^\dagger&\hat{S}^\dagger&\hat{P}+\hat{Q}&\sqrt{2}\hat{R}^\dagger&\frac{1}{\sqrt{2}}\hat{S}^\dagger\\
\frac{1}{\sqrt{2}}\hat{S}^\dagger&\sqrt{2}\hat{Q}&-\sqrt{\frac{3}{2}}\hat{S}&\sqrt{2}\hat{R}&\hat{P}+\Delta_\mathrm{so}&0\\
-\sqrt{2}\hat{R}^\dagger&-\sqrt{\frac{3}{2}}\hat{S}^\dagger&-\sqrt{2}\hat{Q}&\frac{1}{\sqrt{2}}\hat{S}&0&\hat{P}+\Delta_\mathrm{so}
\end{bmatrix} + V_{QD},
\end{equation}
where
\begin{subequations}
\label{eq:LK_terms}
\begin{align}
\hat{P}&=\frac{\hbar^2}{2m_e}\gamma_1\left(\hat{k}_x^2+\hat{k}_y^2+\hat{k}_z^2\right)+\Delta_Px_\mathrm{Si}, \label{eq:P_defn} \\
\hat{Q}&=\frac{\hbar^2}{2m_e}\gamma_2\left(\hat{k}_x^2+\hat{k}_y^2-2\hat{k}_z^2\right)+\Delta_Qx_\mathrm{Si}, \label{eq:Q_defn} \\
\hat{R}&=\frac{\hbar^2}{2m_e}\sqrt{3}\left(-\gamma_3\left(\hat{k}_x^2-\hat{k}_y^2\right)+i\gamma_2\left(\hat{k}_x\hat{k}_y+\hat{k}_y\hat{k}_x\right)\right),~\text{and} \\
\hat{S}&=\frac{\hbar^2}{2m_e}\sqrt{3}\gamma_3\left(\hat{k}_x\hat{k}_z+\hat{k}_z\hat{k}_x-i\hat{k}_y\hat{k}_z-i\hat{k}_z\hat{k}_y\right).
\end{align}
\end{subequations}
$\gamma_1$, $\gamma_2$, and $\gamma_3$ are the Luttinger parameters, $\Delta_\mathrm{so}$ is the offset of the split-off band, and the electrostatic confining potential, $V_{QD}$, acts strictly along the diagonal.
Our 4-band model simply decouples the last two rows/columns corresponding to the split-off band (i.e., we consider only the upper-left 4$\times$4 block).
Terms in $\hat{P}$ and $\hat{Q}$ account for biaxial strain due to the lattice mismatch between the Ge and SiGe layers, where $x_\mathrm{Si}$ is the Si fraction in the SiGe layer, $\Delta_P=-101.4$ meV, and $\Delta_Q=-151.6$ meV (see Appendix~\ref{app:strain_and_band_offsets} for a derivation of these values).
The momentum operators account for coupling to the magnetic field and are expressed in the position basis as
\begin{equation}
\hat{k}_\mu=-i\frac{\partial}{\partial\mu}+\frac{q_e}{\hbar}A_\mu,
\end{equation}
where $A_\mu$ is the $\mu$ component of the associated magnetic vector potential and $q_e$ is the fundamental electron charge.
Note that since we use the convention of adding $+\frac{q_e}{\hbar}A_\mu$ with our Peierls substitution, our Zeeman-like term projected down to the HH and LH spaces is
\begin{equation}
    \hat{H}_Z=+\mu_B\hat{\pmb{J}}\cdot\pmb{B},
\end{equation}
rather than having a minus sign, where $\hat{\pmb{J}}$ is a vector of the operators we have defined as
\begin{equation}
    \hat{J}_x=\begin{bmatrix}
        0&\sqrt{3}\kappa+\frac{7\sqrt{3}}{4}q&0&\frac{3}{2}q&-\sqrt{\frac{3}{2}}(1+\kappa)&0\\
        \sqrt{3}\kappa+\frac{7\sqrt{3}}{4}q&0&2\kappa+5q&0&0&-\frac{1+\kappa}{\sqrt{2}}\\
        0&2\kappa+5q&0&\sqrt{3}\kappa+\frac{7\sqrt{3}}{4}q&\frac{1+\kappa}{\sqrt{2}}&0\\
        \frac{3}{2}q&0&\sqrt{3}\kappa+\frac{7\sqrt{3}}{4}q&0&0&\sqrt{\frac{3}{2}}(1+\kappa)\\
        -\sqrt{\frac{3}{2}}(1+\kappa)&0&\frac{1+\kappa}{\sqrt{2}}&0&0&1+2\kappa\\
        0&-\frac{1+\kappa}{\sqrt{2}}&0&\sqrt{\frac{3}{2}}(1+\kappa)&1+2\kappa&0
    \end{bmatrix},
\end{equation}
\begin{equation}
    \hat{J}_y=i\begin{bmatrix}
        0&-\sqrt{3}\kappa-\frac{7\sqrt{3}}{4}q&0&\frac{3}{2}q&\sqrt{\frac{3}{2}}(1+\kappa)&0\\
        \sqrt{3}\kappa+\frac{7\sqrt{3}}{4}q&0&-2\kappa-5q&0&0&\frac{1+\kappa}{\sqrt{2}}\\
        0&2\kappa+5q&0&-\sqrt{3}\kappa-\frac{7\sqrt{3}}{4}q&-\frac{1+\kappa}{\sqrt{2}}&0\\
        -\frac{3}{2}q&0&\sqrt{3}\kappa+\frac{7\sqrt{3}}{4}q&0&0&-\sqrt{\frac{3}{2}}(1+\kappa)\\
        -\sqrt{\frac{3}{2}}(1+\kappa)&0&\frac{1+\kappa}{\sqrt{2}}&0&0&-1-2\kappa\\
        0&-\frac{1+\kappa}{\sqrt{2}}&0&\sqrt{\frac{3}{2}}(1+\kappa)&1+2\kappa&0
    \end{bmatrix},
\end{equation}
and
\begin{equation}
    \hat{J}_z=\begin{bmatrix}
        3\kappa+\frac{27}{4}q&0&0&0&0&0\\0&\kappa+\frac{1}{4}q&0&0&\sqrt{2}(1+\kappa)&0\\0&0&-\kappa-\frac{1}{4}q&0&0&\sqrt{2}(1+\kappa)\\0&0&0&-3\kappa-\frac{27}{4}q&0&0\\
        0&\sqrt{2}(1+\kappa)&0&0&1+2\kappa&0\\0&0&\sqrt{2}(1+\kappa)&0&0&-1-2\kappa
    \end{bmatrix},
\end{equation}
and $\kappa=3.41$ and $q=0.06$ are additional material constants for Ge.

\subsection{Numerical values for the strain and band offsets}
\label{app:strain_and_band_offsets}
We briefly derive values for the $\Delta_P$ and $\Delta_Q$ coefficients.
In a Ge/SiGe heterostructure, the Ge lattice is typically assumed to be axially strained due to the lattice mismatch with the SiGe layers surrounding it.
We can approximate the lattice spacing in the $xy$-plane of the Ge layer as being equal to the effective lattice constant in the SiGe layer.
The induced strain can be expressed as
\begin{equation}
\epsilon_{xx}=\epsilon_{yy}=\left(1-\frac{a_\mathrm{Si}}{a_\mathrm{Ge}}\right)x_\mathrm{Si},
\end{equation}
where $a_\mathrm{X}$ is the equilibrium lattice constant of material $\mathrm{X}$ and $\epsilon_{xx}$ ($\epsilon_{yy}$) is the strain along the $x$-axis ($y$-axis).
The strain along $z$, $\epsilon_{zz}$, can be calculated from the equilibrium strain tensor as
\begin{equation}
\epsilon_{zz}=-\frac{2C_{12}}{C_{11}}\epsilon_{xx},
\end{equation}
where $C_{11}=129.2$ GPa and $C_{12}=47.9$ GPa are the elastic constants for Ge.
We assume the shear strain (i.e., off-diagonal elements of $\epsilon_{\mu\nu}$) to be zero in our heterostructure, though this might not be the case for other device designs (e.g., those based on hut wires).

Making use of the Bir-Pikus Hamiltonian, this strain has two effects.
The first is to shift the energy levels of all of the hole states in the Ge layer through a linear correction to $\hat{P}$ in Eq.~\ref{eq:P_defn},
\begin{equation}
P_\epsilon=-a_v\left(\epsilon_{xx}+\epsilon_{yy}+\epsilon_{zz}\right)
\end{equation}
where $a_v=2$ eV is a deformation potential constant for Ge.
The second is to change the spacing between the LH and HH bands and their coupling to the SO band, this time through a linear correction to $\hat{Q}$ in Eq.~\ref{eq:Q_defn},
\begin{equation}
Q_\epsilon=-\frac{b_v}{2}\left(\epsilon_{xx}+\epsilon_{yy}-2\epsilon_{zz}\right)
\end{equation}
where $b_v=2.16$ eV is another deformation potential constant.
These contributions can be simplified to yield forms for corrections to $\hat{P}$ and $\hat{Q}$ linear in $x_\mathrm{Si}$ with the previously indicated values for $\Delta_P$ and $\Delta_Q$.
Shear strain is not considered in this work.

We also briefly consider the value of the relevant band offset between the Ge and SiGe layers, which is also linear in $x_\mathrm{Si}$. 
For a pure unstrained Ge well, the offset to a SiGe alloy\cite{schaffler1997high} is
\begin{equation}
    V_0=\Delta_vx_\mathrm{Si},
\end{equation}
where $\Delta_v=-410$ meV.
The strain in $\Delta_Px_\mathrm{Si}$ further adds to this band offset, and for the value of $x_\mathrm{Si}=0.25$ used throughout this work the resulting total offset is $\left(\Delta_v+\Delta_P\right)x_\mathrm{Si}=-128$ meV. 
This value is consistent with density functional theory calculations elsewhere in Ref.~\cite{terrazos2021theory}.

\section{Numerical methods}
\label{app:numerical_methods}

\renewcommand{\theequation}{B\arabic{equation}}
\renewcommand{\thefigure}{B\arabic{figure}}
\setcounter{figure}{0}

\subsection{Discontinuous Galerkin discretization of the Luttinger-Kohn Hamiltonian}
\label{app:dg_discretization}

All results in this manuscript are based on numerical solutions to the LK Hamiltonian in Appendix~\ref{app:extended_LK}.
In particular, we use an interior penalty discretization within a nodal discontinuous Galerkin (DG) framework \cite{hesthaven2007nodal} to render the few-hole problem as a finite-dimensional generalized eigenproblem that is solved using the Locally Optimal Block Preconditioned Conjugate Gradient (LOBPCG) method~\cite{knyazev2001toward,2020SciPy-NMeth}.
The DG framework is used throughout our group's software package, \textsc{Laconic}, which implements these calculations in Python. 
Some of the more salient features specific to this manuscript include ease of treating material discontinuities (see Appendix~\ref{app:discontinuities_in_materials}) and higher-order accuracy.

Like many other PDE solvers (e.g., those based on more conventional finite element or finite volume methods), the DG framework represents the quantum dot in terms of compactly supported basis functions defined on a tessellation of the computational domain $\Omega$ in terms of elements $\mathcal{D}^{(a)}$, such that $\Omega=\cup_a\mathcal{D}^{(a)}$, where $a$ indexes each element.
Each element has an element-local ordinal index $j$ for a set of basis functions that is supported strictly on $\mathcal{D}^{(a)}$ and its boundary $\partial \mathcal{D}^{(a)}$.
For the results in this manuscript, we use hexahedral elements and nodal polynomials complete to some fixed order.
In simpler terms, we decompose the support of the few-hole wave function into nonoverlapping bricks, with a complete set of polynomials defined on each brick that describes the wave function itself.

Unlike more conventional finite element methods, the definition of the basis in DG is such that continuity of the numerical solution is not explicitly guaranteed across mesh elements.
While this facilitates the treatment of discontinuities at material interfaces in the heterostructure, it also allows for discontinuous solutions to the LK Hamiltonian.
While we want to allow for the coefficients of the Hamiltonian to be discontinuous, the associated wave function should still be continuous.
Discontinuities in the numerical wave function are systematically controlled using an interior penalty method that approximately enforces continuity at inter-element interfaces~\cite{arnold1982interior}.
While we will not describe the adaptation of this method to Eq.~\ref{eq:LK-Hamiltonian} in exhaustive detail, we briefly outline the modifications required to ensure a consistent discretization of the Laplacian. 
This can then be generalized to the rest of Eq.~\ref{eq:LK-Hamiltonian}.

Thanks to the element-local support of the basis functions, the matrix corresponding to the discretization of any local linear operator will be divided into blocks corresponding to basis functions on the same element (diagonal blocks) or basis functions on neighboring elements (off-diagonal blocks).
For the Laplacian, the matrix elements associated with the $j$th and $k$th basis functions on $\mathcal{D}^{(a)}$ are 
\begin{align}
\mathcal{L}_{jk}^{(a)}=&\int_{\mathcal{D}^{(a)}}\!d^3\pmb{r}\,\nabla\ell_j^{(a)}\cdot\nabla\ell_k^{(a)}-\frac{1}{2}\oint_{\partial\mathcal{D}^{(a)}}\!d^2\pmb{r}\,\left(\ell_j^{(a)}\vec{n}^{(a)}\cdot\nabla\ell_k^{(a)}+\ell_k^{(a)}\vec{n}^{(a)}\cdot\nabla\ell_j^{(a)}\right)+\tau\oint_{\partial\mathcal{D}^{(a)}}\!d^2\pmb{r}\,\ell_j^{(a)}\ell_k^{(a)},
\end{align}
where the one may interpret the first term as arising due to integration by parts and the second and third terms account for the multi-valued nature of the function space spanned by our basis.
The matrix elements associated with the $j$th basis function on $\mathcal{D}^{(a)}$ and the $k$th basis function on $\mathcal{D}^{(b)}$ are
\begin{align}
\mathcal{L}_{jk}^{(a,b)}=&\frac{1}{2}\int_{\partial\mathcal{D}^{(a)}\cap\partial\mathcal{D}^{(b)}}\!d^2\pmb{r}\,\left(\ell_j^{(a)}\vec{n}^{(a)}\cdot\nabla\ell_k^{(b)}+\ell_k^{(b)}\vec{n}^{(a)}\cdot\nabla\ell_j^{(a)}\right)-\tau\int_{\partial\mathcal{D}^{(a)}\cap\partial\mathcal{D}^{(b)}}\!d^2\pmb{r}\,\ell_j^{(a)}\ell_k^{(b)},
\end{align}
mirroring the form of the diagonal blocks.
We note that the discretized Laplacian is manifestly Hermitian.
The positive real number $\tau$ is the interior penalty parameter that enforces the approximate continuity of eigenstates from one mesh element to the next.
Larger values of $\tau$ more precisely enforce continuity of the eigenfunctions at the expense of increasing the condition number of the associated matrix.
One can also think of $\tau$ as being related to an energetic penalty associated with unphysical discontinuous eigenfunctions.
Our implementation makes use of a simple heuristic for $\tau$ that accounts for variation in the sizes of the mesh elements and the order of local expansions on them, such that ill conditioning is rarely an issue for the low-energy subspace of eigenvectors with which we are primarily concerned.
Other operators in Eq.~\ref{eq:LK-Hamiltonian} are discretized similarly.

\subsection{Discontinuities in material parameters}
\label{app:discontinuities_in_materials}
We consider the effects of a finite potential barrier at the interfaces of the Ge layer by explicitly including the surrounding SiGe layers in our computational domain $\Omega$, rather than enforcing a homogeneous Dirichlet boundary condition at the boundary of the Ge layer (which is equivalent to an infinite potential barrier).
However, the Luttinger parameters differ between the Ge and SiGe layers, leading to discontinuities in quantities (coefficients in Eq.~\ref{eq:LK-Hamiltonian}) that are typically spatially uniform (e.g., the effective masses).
An additional derivative interior penalty term is included in our discretization to properly satisfy the resulting Ben-Daniel-Duke boundary conditions~\cite{wang2022modelling}.
The matrix elements associated with this term are
\begin{equation}
\mathcal{C}_{jk}^{(a,b)}=\tau_d\int_{\partial\mathcal{D}^{(a)}\cap\partial\mathcal{D}^{(b)}}\!d^2\pmb{r}\,\left(\vec{n}^{(a)}\cdot\hat{m}_\mathrm{eff}^{-1}\nabla\ell_j^{(a)}\right)\left(\vec{n}^{(b)}\cdot\hat{m}_\mathrm{eff}^{-1}\nabla\ell_k^{(b)}\right)
\end{equation}
for both on- (a and b are the same element) and off-diagonal blocks (a and b are neighboring elements). 
Here $\hat{m}_\mathrm{eff}^{-1}$ is the inverse of the effective mass matrix, which can be derived from Eq.~\ref{eq:LK-Hamiltonian}, and $\tau_d$ is a positive real-valued derivative interior penalty parameter similar to $\tau$ in Appendix~\ref{app:dg_discretization}.
This derivative interior penalty term increases the energy of numerical solutions that do not satisfy the Ben-Daniel-Duke boundary conditions, leading to solutions that do satisfy these conditions~\cite{brickson2023modeling}.
Like $\tau$, $\tau_d$ is set according to a simple heuristic that ensures that the low-energy states of interest satisfy this boundary condition.

\subsection{Hartree-Fock and Configuration Interaction implementations}
\label{app:HF_and_CI}
To capture the effect of multi-hole interactions, we use a Configuration Interaction (CI) expansion based on an active space derived from a self-consistent Hartree-Fock calculation~\cite{szabo2012modern,gamble2021advanced}.
For the Hartree-Fock calculation we define a sequence of operators, $\hat{H}_\mathrm{HF}^{(n)}$ with $\hat{J}_\alpha^{(n)}$ and $\hat{K}_\alpha^{(n)}$, such that
\begin{equation}
\hat{H}_\mathrm{HF}^{(n)}=\hat{H}_{LK}+\sum_{\alpha=0}^{N-1}\left[\hat{J}_\alpha^{(n)}-\hat{K}_\alpha^{(n)}\right]
\end{equation}
where $\hat{H}_{LK}$ is the single-particle Luttinger-Kohn Hamiltonian (see Eq.~\ref{eq:LK-Hamiltonian}), $N$ is the number of holes in the system, and $\hat{J}_\alpha^{(n)}$ and $\hat{K}_\alpha^{(n)}$ are the Hartree and exchange operators, respectively.
The index $n$ is associated with the $(n+1)$th step of a self-consistent iteration with initial condition
\begin{equation}
\hat{J}_\alpha^{(0)}=\hat{K}_\alpha^{(0)}=0
\end{equation}
and subsequent iterates
\begin{equation}
\hat{J}_\alpha^{(n+1)}\phi(\pmb{r}_1,\sigma_1)=\mathcal{V}\left\{\left|\psi_\alpha^{(n)}(\pmb{r}_2,\sigma_2)\right|^2\right\}(\pmb{r}_1)\phi(\pmb{r}_1,\sigma_1) \label{eq:hartree_defn}
\end{equation}
and
\begin{equation}
\hat{K}_\alpha^{(n+1)}\phi(\pmb{r}_1,\sigma_1)=\mathcal{V}\left\{{\psi_\alpha^{(n)}}^*(\pmb{r}_2,\sigma_2)\phi(\pmb{r}_2,\sigma_2)\right\}(\pmb{r}_1)\psi_\alpha^{(n)}(\pmb{r}_1,\sigma_1) \label{eq:exchange_defn},
\end{equation}
with $\mathcal{V}$ defined as the solution to 
\begin{equation}
-\nabla\cdot\left(\epsilon(\pmb{r})\nabla\mathcal{V}\left\{f(\pmb{r},\sigma)\right\}(\pmb{r})\right)=\sum_\sigma f(\pmb{r},\sigma) \label{eq:poisson}
\end{equation}
where $\epsilon(\pmb{r})$ is the material-dependent electric permittivity and $f$ takes the forms evident in the braces in Eq.~\ref{eq:hartree_defn} or Eq.~\ref{eq:exchange_defn} depending on whether $J_{\alpha}^{(n)}$ or $K_{\alpha}^{(n)}$ are being computed.
Note that $\alpha$ indexes eigenstates of $\hat{H}_\mathrm{HF}^{(n)}$ in both space and spin, which are mixed due to SOC.
For computational efficiency, we have replaced the Coulomb integrals that typically define $J_{\alpha}^{(n)}$ and $K_{\alpha}^{(n)}$ with solutions to a Poisson equation (Eq.~\ref{eq:poisson}) that is is discretized using an interior penalty method similar to the one described in Appendix~\ref{app:dg_discretization}.
The resulting linear system of equations is solved iteratively.
Self-consistency of the Hartree-Fock calculation is achieved when the occupied orbital energies, $E_\alpha^{(n)}$, satisfy
\begin{equation}
\left|E_\alpha^{(n+1)}-E_\alpha^{(n)}\right|<\eta,\,\,\,\forall\alpha<N,
\end{equation}
where $\eta$ is a convergence parameter that is set to $1~\mu$eV for the calculations in this manuscript.

Eigenstates of the self-consistent Hartree-Fock operator $\{\psi_\alpha(\pmb{r},\sigma)\}_\alpha$ are then used to generate the matrix of interactions for a CI calculation,
\begin{equation}
U_{\alpha\beta\gamma\delta}=\int\!d^3\pmb{r}_1\,\mathcal{V}\left\{\psi_\alpha^*(\pmb{r}_2,\sigma_2)\psi_\gamma(\pmb{r}_2,\sigma_2)\right\}(\pmb{r}_1)\sum_{\sigma_1}\psi_\beta^*(\pmb{r}_1,\sigma_1)\psi_\delta(\pmb{r}_1,\sigma_1).
\end{equation}
With this and the single-particle matrix
\begin{equation}
h_{\alpha\beta}=\sum_\sigma\int\!d^3\pmb{r}\,\psi_\alpha^*(\pmb{r},\sigma)\hat{H}_{LK}\psi_\beta(\pmb{r},\sigma)
\end{equation}
the full CI Hamiltonian takes the form
\begin{equation}
\hat{H}_\mathrm{CI}=\sum_{\alpha,\beta}h_{\alpha\beta}\hat{c}_\alpha^\dagger\hat{c}_\beta+\sum_{\alpha,\beta>\alpha,\gamma,\delta>\gamma}U_{\alpha\beta\gamma\delta}\hat{c}_\alpha^\dagger\hat{c}_\beta^\dagger\hat{c}_\gamma\hat{c}_\delta
\end{equation}
where $\hat{c}_\alpha^\dagger$ ($\hat{c}_\alpha$) is a creation (annihilation) operator for single-particle orbital $\alpha$.
This matrix is computed over a subspace with fixed particle number and diagonalized to yield the low-lying multi-hole wave functions for that particle number.
Note that not all $N$-particle CI Hamiltonians were built with states from $N$-particle HF calculations.
This is due to a few factors.
First, the change in the energy levels with respect to magnetic field was included as a perturbative addition in the CI matrices to accomodate calculating these energies at a large number of field values more efficiently.
Second, because we want $g$ factors from our calculations, capturing the gap between energy levels accurately is more important in this context than capturing the ground state energy.
The exchange operator in the HF calculation reduces the energy of the states higher than $N$ that match the first $N$ in total spin more than a state with opposite spin, which can be easily shown for a 3-particle state with no magnetic field.
As such, we end up using CI Hamiltonians from HF calculations of fewer states in order to avoid artificially widening the Zeeman gap, e.g. using 2-particle HF calculations to generate the states for 3-particle CI Hamiltonians.

\subsection{Image charges for Poisson solves}
\label{app:image_charges}
In Appendix~\ref{app:HF_and_CI} we noted that the material-dependent electric permittivity $\epsilon(\pmb{r})$ is incorporated to screen the Hartree and exchange interactions. 
This is due to the fact that we are relying on a low-energy (Luttinger-Kohn) effective description of the holes, rather than a strict first principles model.
However, this screening model requires that we consistently account for the fact that there are differences in the permittivities of the Ge and SiGe layers in Eq.~\ref{eq:poisson}.
In short, the effective Coulomb interaction $\mathcal{V}$ needs to account for the self-energy of any given hole's interaction with its images in the adjacent layers, as well as images of other holes.
To do so, we use a truncated set of image charges.

For example, with a charge in the Ge layer, satisfying the standard electrostatic boundary conditions at the interface with the top SiGe layer induces an image charge in that layer with an effective charge proportional to
\begin{equation}
q_\mathrm{image}\propto\frac{\epsilon_\mathrm{Ge}-\epsilon_\mathrm{SiGe}}{\epsilon_\mathrm{Ge}+\epsilon_\mathrm{SiGe}}.
\end{equation}
To then satisfy the boundary conditions at the interface with the bottom SiGe layer, we get a similar image charge as the one above and another one responding to that first image charge with
\begin{equation}
q_\mathrm{image}'\propto\left(\frac{\epsilon_\mathrm{Ge}-\epsilon_\mathrm{SiGe}}{\epsilon_\mathrm{Ge}+\epsilon_\mathrm{SiGe}}\right)^2.
\end{equation}
This series of image charges goes on infinitely, but the exponential suppression of image charges of image charges (and so on) continues such that, for our particular material system the charge of the $n$th image is $\propto1/31^n$.
Thus we truncate the series at $n=1$.

\section{Maximum-likelihood estimation of $V_{QD}$ parameters}
\label{app:mle_details}

\renewcommand{\theequation}{C\arabic{equation}}
\renewcommand{\thefigure}{C\arabic{figure}}
\setcounter{figure}{0}

\subsection{Handling of experimental data}
\label{app:handling_data}

Our results are based on the raw experimental data that were first presented in Ref.~\cite{miller2022effective}.
These data consist of the logarithmic derivative of the current through a charge sensor as a function of the applied magnetic field strength and the voltage controlling the dot occupation.
The charge transitions of interest are identified with the peaks of this logarithmic derivative and the voltages are converted to energies through a measured lever arm.
Ten separate scans were taken and we partially compensate for device drift by considering energy differences.

The average charging energy from $N$ to $N+1$ at magnetic field strength $B_z$ is estimated as the average over all scans within a small range $B_z+\pm\delta B_z$ for $\delta B_z=\SI{5}{\milli\tesla}$, where the fields at which the measurements were taken are labeled as $\{B_m\}_m$ and the attendant charging energy estimates are $\lambda(B_m)$.
The same sample is used to estimate the standard deviation, $\sigma(B_m)$, used in our likelihood calculations.
To account for the fact that some scans captured data over a larger range of magnetic field strengths than others, the log-likelihood is weighted by the number of data points used to estimate any given mean. 
The final log-likelihood is normalized by the total number of data points.

\subsection{Maximum-likelihood formulas}

We used the final form of our LK Hamiltonian (6 bands with finite confinement and CI) to calculate the charge transition energies of a QD as a function of magnetic field and QD potential parameters, $\tilde{\lambda}(B;V_{QD}\left[l_x,l_y\right])$.
Here $V_{QD}$ is of the form
\begin{equation}
    V_{QD}\left[l_x,l_y\right]=\frac{m_e}{2\left(\gamma_1+\gamma_2\right)}\left(\omega_x^2x^2+\omega_y^2y^2\right)+E_{z}z=\frac{\left(\gamma_1+\gamma_2\right)\hbar^2}{2m_e}\left(\frac{x^2}{\ell_x^4}+\frac{y^2}{\ell_y^4}\right)+E_{z}z,
\end{equation}
where $m_e$ is the rest mass of a free electron, $\omega_\mu$ is the oscillator strength along axis $\mu$ consistent with a ground state Gaussian wavefunction of standard deviation $\ell_\mu$, and $E_z$ is the electric field strength.
We consider dot widths in the interval $\ell_y\in[70~\text{nm},130~\text{nm}]$ with anisotropies in the interval $(\omega_x-\omega_y)/\hbar\in[0~\text{meV},1~\text{meV}]$.
We consider a fixed value of $E_z=1.5~\text{V/$\mu$m}$ to both reduce the dimensionality of the inference and to account for the fact that the vertical field was constrained in other simulations supporting the design of the device in Ref.~\cite{miller2022effective}. 

The likelihood of $\tilde{\lambda}$ is
\begin{equation}
    L\left(\left\{\left(\lambda(B_m),
    \sigma(B_m)\right)\right\}_m,\tilde{\lambda}(B;V)\right)=\prod_m\frac{e^{-\left(\tilde{\lambda}(B_m;V)-\lambda(B_m)\right)^2/2\sigma(B_m)^2}}{\sigma(B_m)\sqrt{2\pi}},
\end{equation}
where it assumed that the measurement error is Gaussian.
The parametrization of $V_{QD}$ that maximizes this quantity is the one most consistent with the data.
To facilitate calculations, we use the logarithm of the likelihood instead of the likelihood itself,
\begin{equation}
    \mathrm{log-L}\left(\left\{\left(\lambda(B_m),
    \sigma(B_m)\right)\right\}_m,\tilde{\lambda}(B;V)\right)=-\sum_m\ln\left(\sigma(B_m)\sqrt{2\pi}\right)-\sum_m\frac{\left(\tilde{\lambda}(B_m;V)-\lambda(B_m)\right)^2}{2\sigma(B_m)^2}.
\end{equation}
The results of this analysis are reported in Fig.~\ref{fig:mle_results} and the basis for our assignment of the shape of our quantum dot.

\subsection{Comparing suboptimal magnetospectroscopy predictions to data}

\begin{figure*}[ht]
\includegraphics{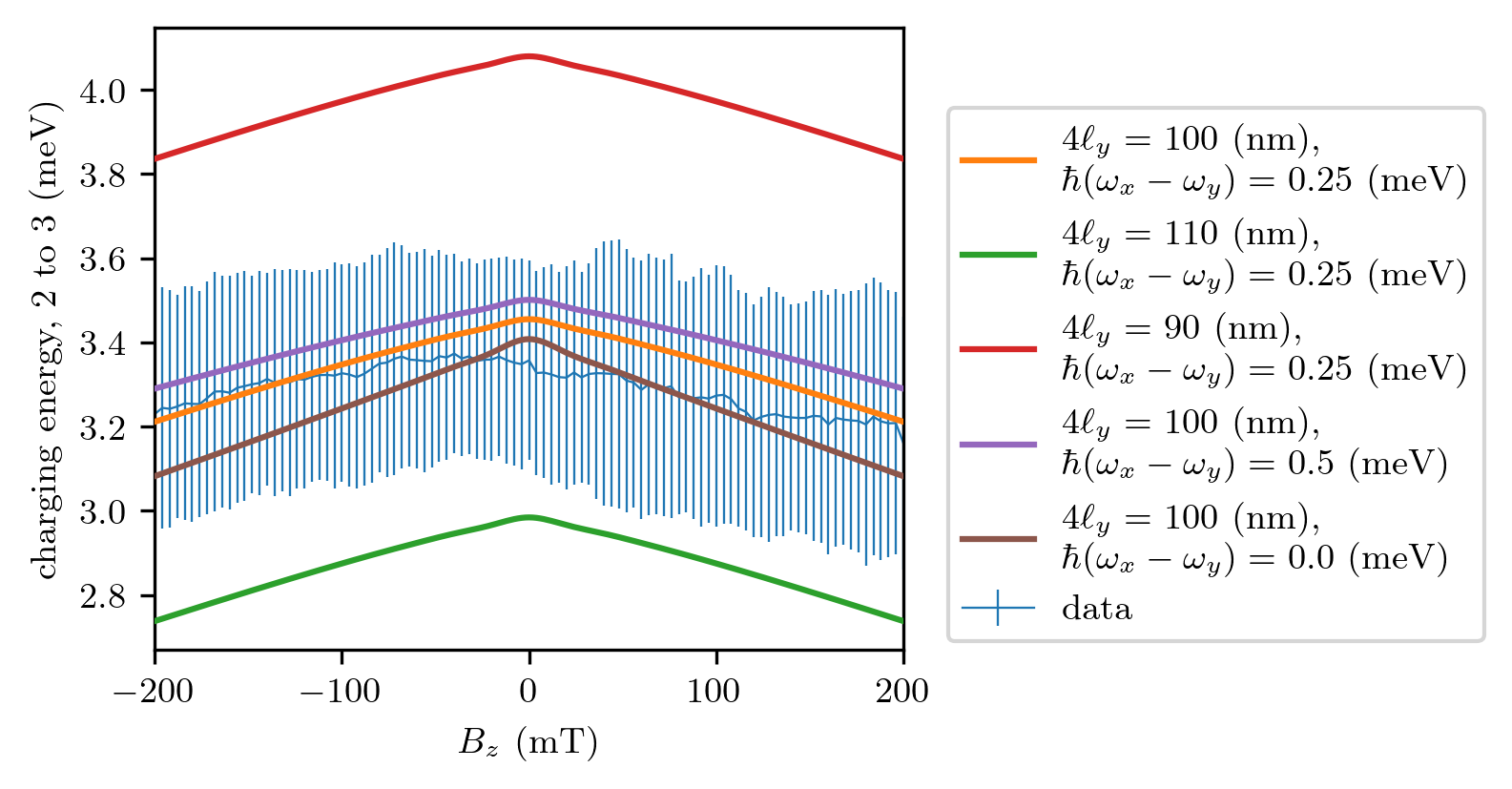}
\includegraphics{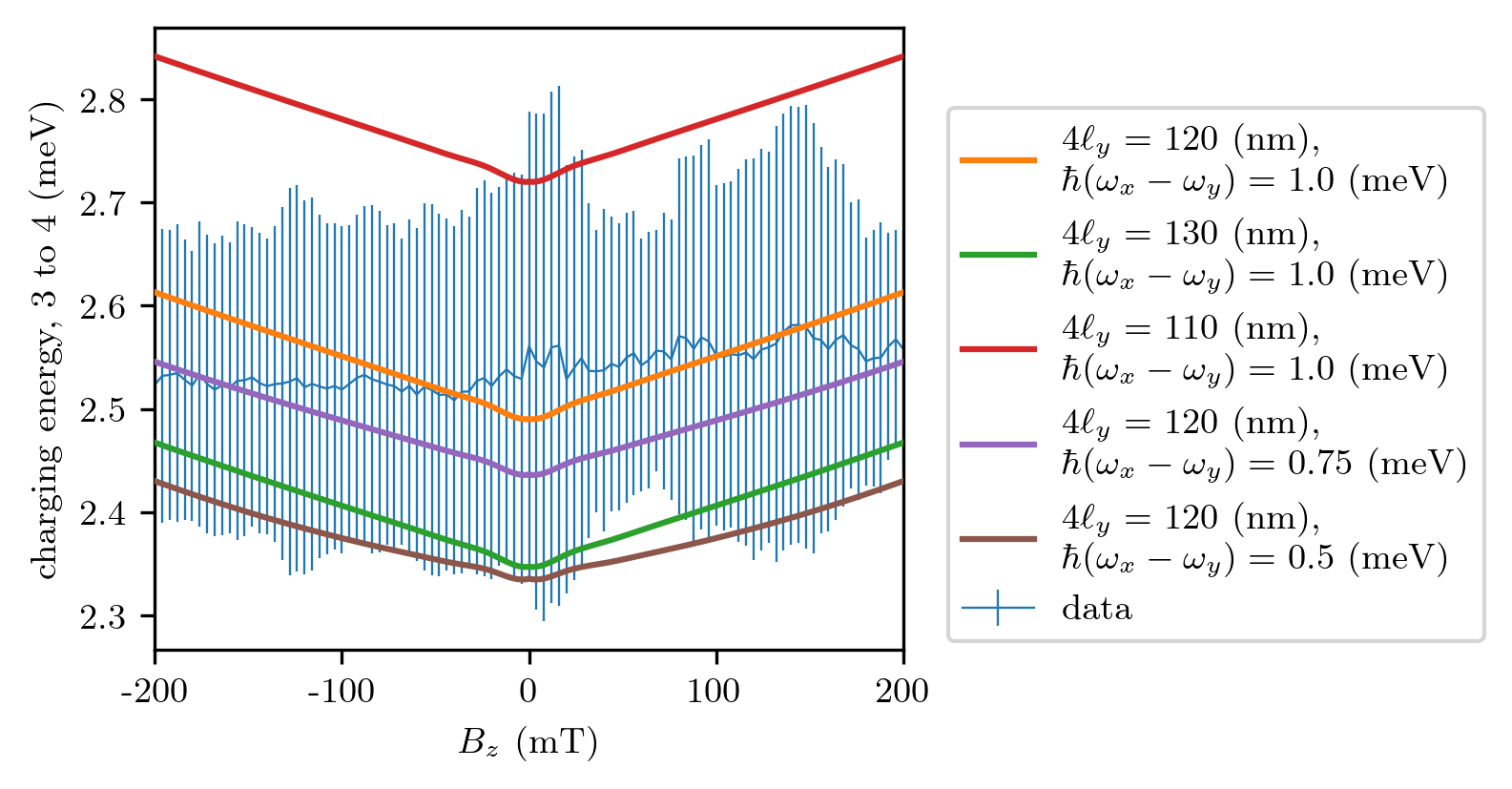}
\caption{
Variations in the predicted QD spectra due to changes in the parameters of $V_{QD}$, compared to the experimental data.
We include the spectra for parameters that achieve the maximum likelihood (orange) as well as small variations about them (all other colors) to visually illustrate the sensitivities of our inference.
\label{fig:SM-suboptimal_MLE_predictions}
}
\end{figure*}

In Fig.~\ref{fig:mle_results}, the ranges of the log-likelihoods calculated suggest that the consistency of the predictions from our model with the experimental data drop off exponentially with the change in the width of the potential.
In Fig.~\ref{fig:SM-suboptimal_MLE_predictions}, we present a direct visual comparison of the predictions and the data for multiple different potentials in addition to the prediction closest to the data to show the sensitivity of the predicted spectra to relatively small changes in the model parameters.
It worth noting that much of the sensitivity in the value of the log-likelihood is due to the change in the charging energy with $V_{QD}$, more than the change in the slopes.
Thus, differentiating between potentials that do a better job of matching the slopes in the magnetospectroscopy data is more difficult for the data at hand and this is compounded by the computational cost of multi-hole calculations.
Future analyses of this sort could be made more efficient by heuristically pre-screening potentials that are more likely to fit the charging energy without doing a full CI calculation.

\section{Zero-field orbital splitting from cubic spin-orbit coupling}
\label{app:zero-field}

We remark briefly on the non-monotonic dependence of the $g$ factor on the QD width, evident in Fig.~\ref{fig:g-factor_comparison}.
This behavior is a function of both the underlying physics and certain details of how the effective $g$ factor is calculated.
The first reason is that at very low magnetic fields, the second and third excited states that we label as constituting the second shell in this paper have both opposite spin and opposite orbital angular momentum, especially for more isotropic dots.
In other words, there is an energy associated with the spin and orbital angular momentum being aligned or anti-aligned.
At higher magnetic fields, the energy splitting from the alignment between the orbital angular momentum and the magnetic field reorders the states so that the second and third excited states have the same orbital character while still having opposite spins, more in line with a typical shell filling picture.
The second reason is that the calculated $g$ factors for Fig.~(\ref{fig:g-factor_comparison}) are averaged over a range of magnetic fields for the sake of aligning with how the magnetospectroscopy spans a range of magnetic fields.
Essentially, the calculation is
\begin{equation}
    \left\langle g_z^{(2)}(B_z)\right\rangle=\frac{\int_{0^+}^{B_\mathrm{max}}\!dB\,\frac{E_3(B_z)-E_2(B_z)}{\mu_BB_z}}{B_\mathrm{max}},
    \label{eq:g2}
\end{equation}
where $E_j(B_z)$ is the energy of the $j$th excited state as a function of the magnetic field along $z$ and $B_\mathrm{max}=\SI{200}{\milli\tesla}$ is the maximum magnetic field strength in the analysis.
Note that this is a simplification of $g$ factors that condenses them entirely to the splitting between consecutive energy levels, which obfuscates the effects of the orbital response to the magnetic field, as well as this zero-field splitting.
Where the transition between the state associated with $E_3(B_z)$ having the orbital angular momentum aligned or anti-aligned with $E_2(B_z)$ is within the averaging range affects the calculated $g$ factor, and wider dots tend to cross through this transition at lower fields.

The dominant contribution to this zero-field, spin-dependent orbital splitting is not reproduced by the typical 2-band model derived starting from the LH+HH space without the SO bands that is often cited in discussions that describe SOC in Ge hole QDs as cubic~\cite{moriya2014cubic,terrazos2021theory,wang2021optimal,rodriguez2023linear}.
While this cubic SOC may contribute somewhat, this phenomenon persists at $E_z=0$ where the cubic terms are zero, which can be shown by calculation~\cite{wang2021optimal} or by symmetry arguments.
As such, this SOC is fundamentally different from that usually discussed through the lens of a 2-band model, and requires either more bands or a modification to the 2-band model.
Here, we provide calculations that show this phenomenology, and leave the description of this physics in a reduced model for future work.

\begin{figure*}[ht]
\includegraphics{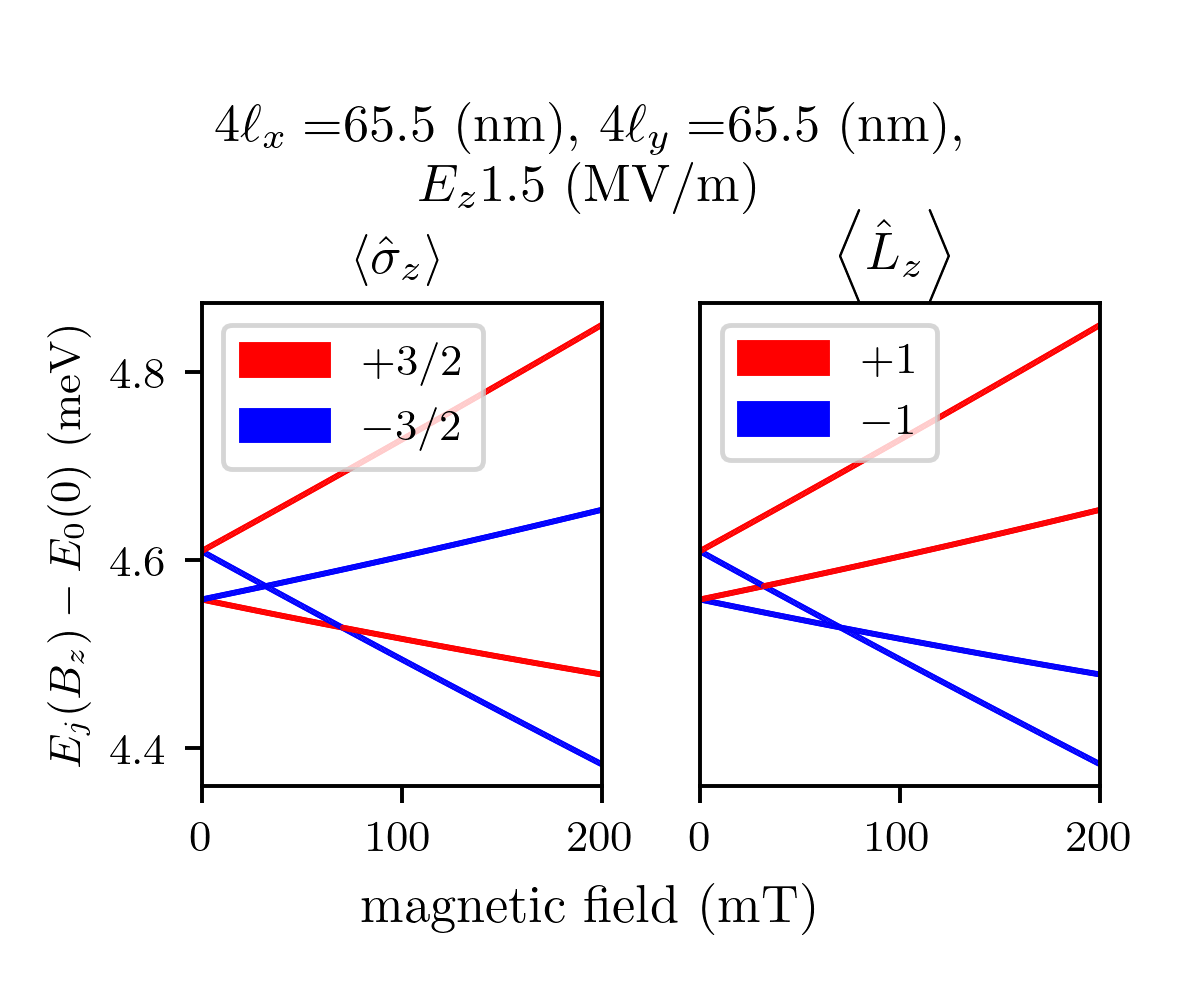}
\includegraphics{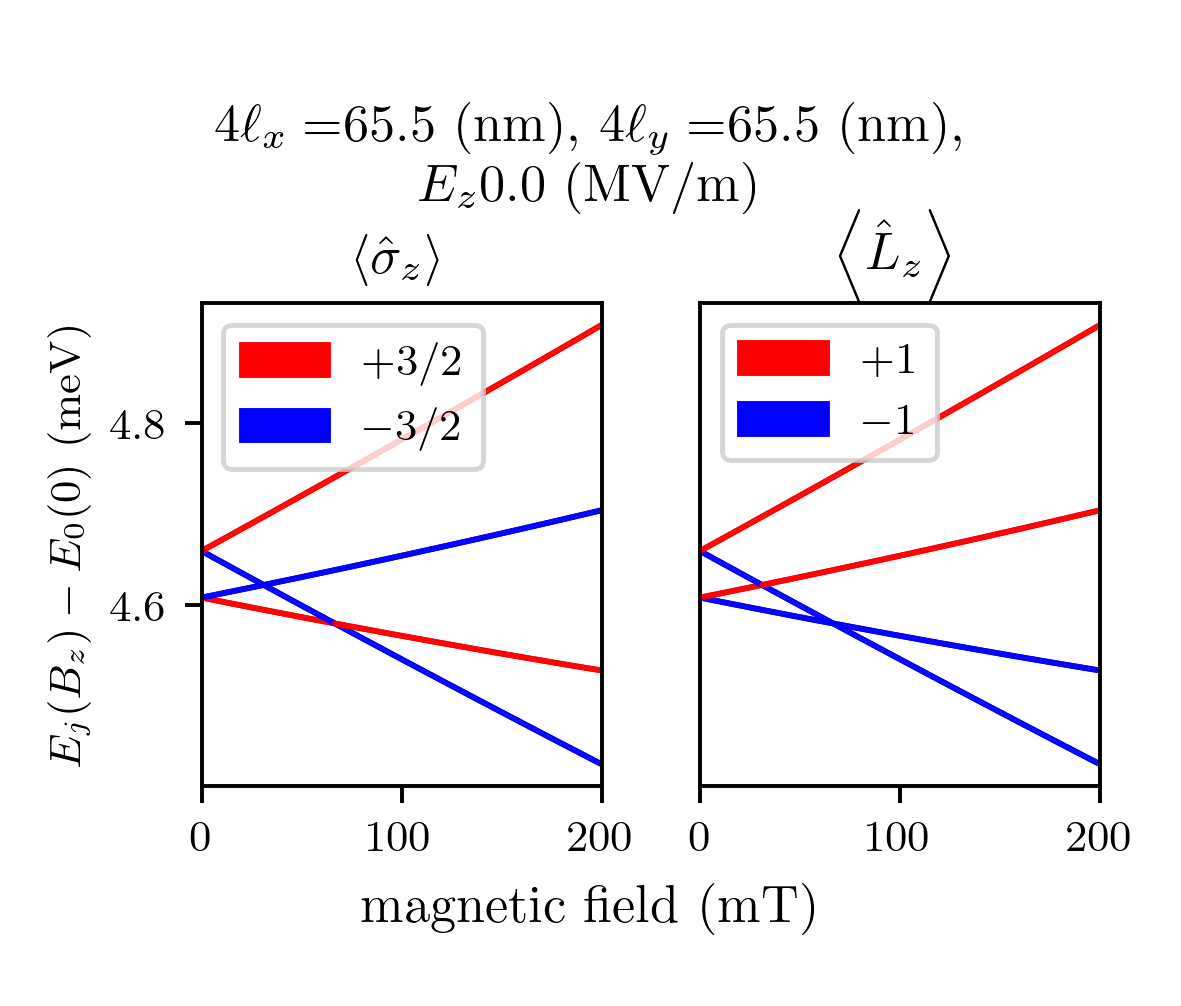}
\includegraphics{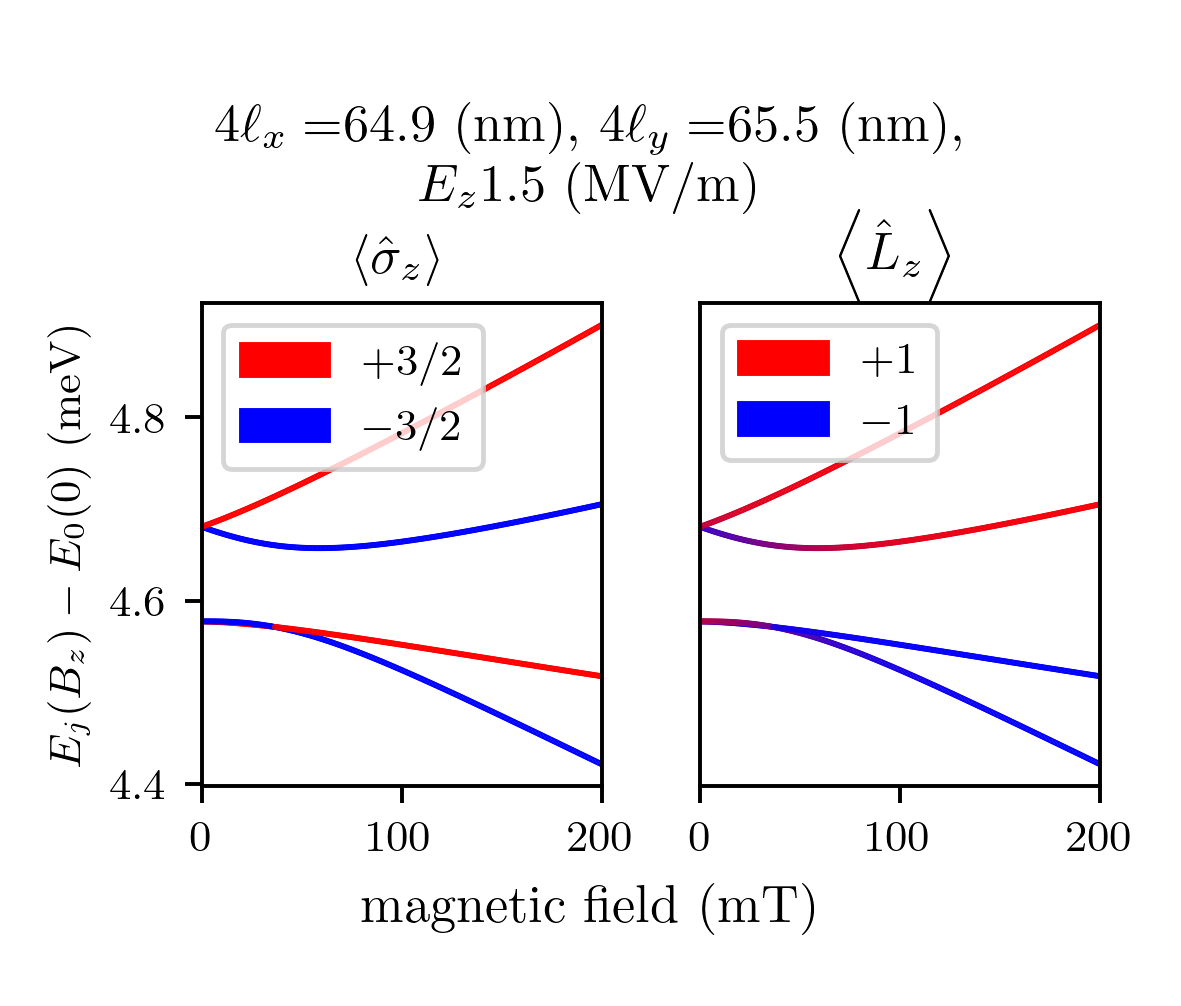}
\includegraphics{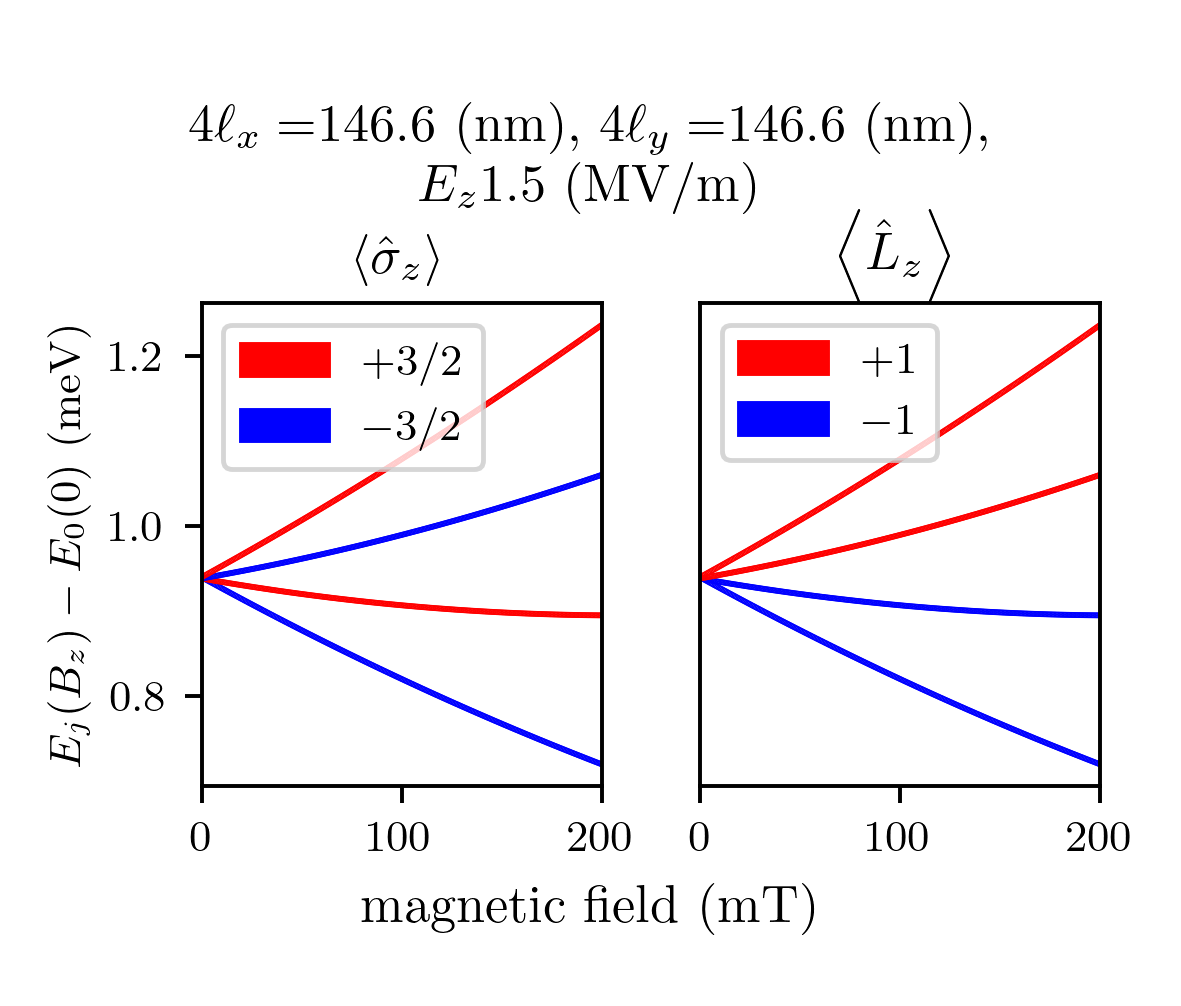}
\caption{
Energy levels as a function of out-of-plane magnetic field strength with spin and orbital angular momentum color coding at different potential widths and electric field strengths.
\label{fig:SM-second_shell_physics}
}
\end{figure*}

To show this, we share plots of the energy spectrum of the second to fifth excited states in Fig.~\ref{fig:SM-second_shell_physics}.
These plots are color coded to reveal the spin and orbital angular momentum alignment as a function of magnetic field strength.
This is shown for various field strengths, including $E_z=0$, and different widths of the potential.
If we consider the second shell to be comprised of the second and third excited states as Eq.~\ref{eq:g2} assumes, the plots show that these states have opposite spin and opposite orbital angular momentum at low field strengths, especially for much smaller dots, and this happens independently of electric field strength.
For very wide dots, this effect is negligible, and the second-shell $g$ factor approaches something similar to the first-shell $g$ factor.
Determining an intuitive explanation for the source of this phenomenon and its effects on qubit operation will be important for understanding higher-occupancy dots in large arrays.

\end{document}